\documentstyle[11pt,paspconf,epsf]{article}

\begin{document}

\title{The NOG Sample: \\
3D Reconstruction of the Real-Space Density Field}

\author{Christian Marinoni, Giuliano Giuricin, Lorenzo Ceriani}

\affil{Department of Astronomy, University of Trieste and Sissa}

\begin{abstract}
We discuss the real-space reconstruction of the 
optical galaxy density field in the local Universe (cz$\leq $6000 km/s)
as derived from
the 7076 galaxies of the Nearby Optical Galaxy (NOG) sample (see 
Giuricin et al. 1999 in the same volume). NOG is  the currently  
best  approximation to a homogeneous all-sky 3D optically selected 
galaxy sample that probes in great detail volumes of cosmological interest.

Our final goal is to construct a reliable, robust and  unbiased
field of density contrasts over a wide range of physical  scales.
 
Exploring in detail the nature of the three dimensional galaxy distribution
will provide  us with invaluable qualitative cosmographical
information about the  topology and morphology of the local overdensities;
but it also allows us to investigate on the  z=0 cosmology,  
greatly increasing our  quantitative understanding of
physical parameters that  constrain the evolution of  
structures and their clustering properties.
Moreover, its near full-sky coverage and the large variety in
galaxy content  make the NOG ideal also for more specific tasks as the  
deconvolution of environmental effects from the properties  and evolution
history of the galaxies.

\end{abstract}

\section{Introduction}
Due to the high resolution with which NOG samples the dominating structures  
of the nearby Universe, this catalog is more suitable than 
IRAS-selected galaxy
samples for mapping the galaxy density field  on quite small scales ($<2$ Mpc).
However, high-density sampling rate, achieved with the NOG selection 
criteria ($B_T^c \leq 14, cz \leq 6000, b \geq |20^{\circ}|$),
is counteracted by systematic effects arising 
from the cutoffs in the selection parameters or non-uniformities in 
the original catalogs which may vave not been properly homogenized in our
sample. So we have tried to correct and  minimize these
biases testing the sample completeness by means of a count-magnitude 
analysis  and deriving the appropriate  
luminosity and redshift selection functions.

Historically, redshift surveys have provided the raw basis for investigating 
the three dimensional nature of the Universe.
However, a large and complete sample of galaxy {\em distances}
would represent a 
marked improvement over redshift surveys for measuring the properties of the 
galaxy 3D distribution since the density map in redshift space can be a 
systematically distorted version of the real picture.
So we have carried out the task of replacing  accurated  ``true distances'' 
measurements for all the galaxies of the sample for which redshift 
information was available.
This has been carried out (Marinoni et al. 1998) by modelling  the Doppler 
perturbations induced by peculiar motions and disentangling 
the cosmological component of the redshift which is the one acting as 
a distance indicator.

\begin{figure}
\plotfiddle{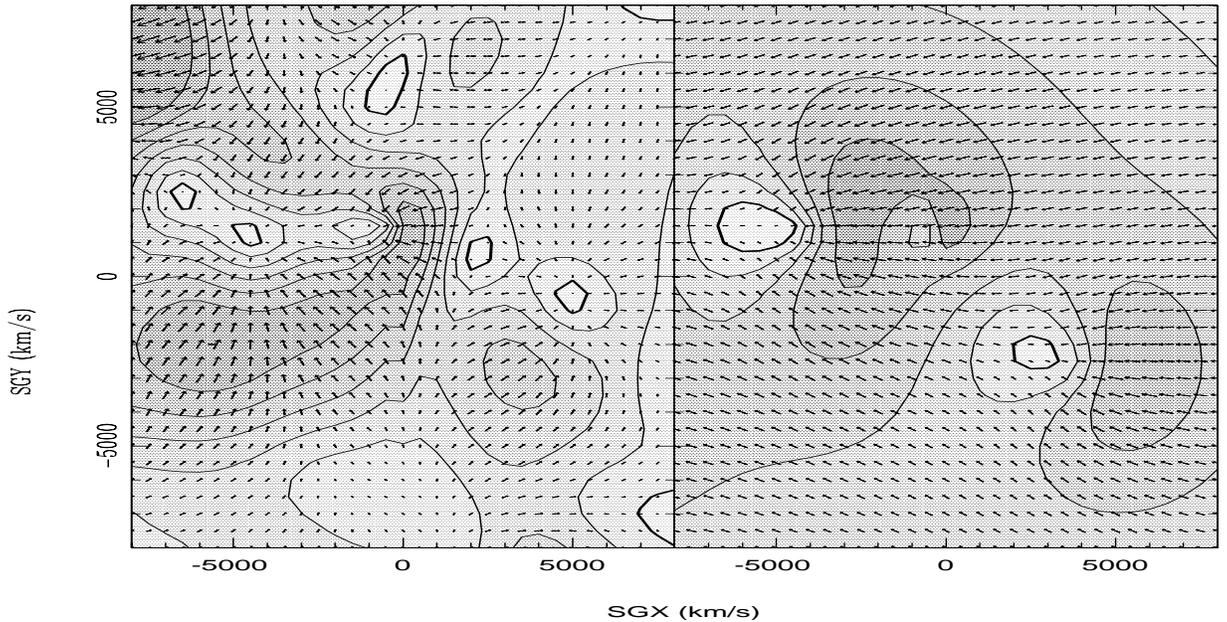}{7.5cm}{0}{85}{42.5}{-268}{-60}

\caption{Plots showing the velocity field in the CMB frame for the 
modified cluster dipole model ({\em left}) and the multi-attractor model 
fitted to the Mark III data set ({\em right}). The vector shown are 
projections of the 3D velocity field in the Supergalactic plane SGX, SGY.
The contours correspond to the same velocity vector modulus; contour spacing
is 100 km $s^{-1}$, the heavy contours marking 100 $km s^{-1}$ and 200
$km s^{-1}$ for the two models respectively.} 
\end{figure}

\section{Distance reconstruction}
In order to correct raw redshift-distances we used two basic models of the 
peculiar velocity field. These two models mean to be  representative of
the two competing and most popular pictures of the the z=0 kinematics. As a matter of fact, they describe the velocity field giving two opposite
interpretations of the amplitude and  the length scale coherence of the 
motions. 
The first model is the optical cluster 3D-dipole reconstruction scheme of 
Branchini \& Plionis (1996) that we modify with the inclusion of a local
model of Virgocentric infall. This model shows a region where the flow 
bifurcates towards the Great Attractor and towards the Perseus-Pisces complex.

The second description of the peculiar velocity field has been worked out
using a multi-attractor model fitted to the Mark III peculiar velocity 
catalogue (Willick et al. 1997). 
This is a collection of homogeneized distances for a sample of
galaxies of different morphological type and distributed in a nearly 
isotropic way in the sky.
In applying this reconstruction  scheme we have adopted
a King density profile for characterizing the mass distribution
of each attractor 
(i.e. Virgo cluster, the Great Attractor, the Perseus-Pisces 
and Shapley superclusters) and a weakly non-linear series expansion
by Reg\"os \& Geller (1989), to relate the peculiar velocity field and the 
mass fluctuations. 
The emerging picture is the one in which the principle feature of the velocity field in the PP region is a coherent streming flows in the general direction 
of the GA and Shapley superclusters.


Inverting the non linear redshift--distance relations predicted by the 
above--mentioned velocity field models, we derive the distances 
of galaxies.  

The use of different velocity field models allows us to check to what 
extent differences in the description of the peculiar flows influences 
the estimate of galaxy distances in the nearby universe.
We note that these differences turn out to be more prominent at the largest 
and smallest distances rather than for intermediate distances (i.e., 
for $2000 < r < 4000$ km/s, where $r$ is the distance expressed in 
km/s).

We have also studied the stability of the luminosity function and
of the derived selection function  against variations in the adopted
peculiar velocity field models. Following the lines described in 
Marinoni et al. 1999, we found that peculiar motion effects are of the order 
of statistical uncertainties and cause at most variations of 1 $\sigma$ in 
$\alpha$ and 2$\sigma$ in $M^{*}_B$.

\section{Density Reconstruction}

The galaxy distribution is intrinsically a point process. 
The problem of reconstructing the density fluctuation $\delta({\bf r})$
is connected with finding the best transformation scheme for {\em diluting}
the point distribution into a continuous density field. After having
devised an algorithm to infer real distances from measured redshifts,
the remaining problems we have to overcome are:
\begin{itemize}
\item  the number density of galaxies,
in a flux-limited redshift sample, is a decreasing
function of distance and a small error in the selection function, used 
to recover the real population of objects, causes
a systematic  error in the density field.
\item  the mean interparticle spacing of a redshift catalog is an
increasing function of distance with corresponding ever increasing
shot noise. Correcting for this effect introduce a lack of statistical
similarity between the nearby and faraway parts of the catalog.
\item there is a 34 \% of the sky which is uncovered by the NOG catalog. 
\end{itemize}

We address all these issues smoothing with a normalized Gaussian filter
having a smoothing length which is a properly defined 
increasing function of distance. Moreover, we assign each galaxy a weight 
given by the inverse of the sample selection function in order to well
calibrate the median value of the density.

The specific features of the galaxy distribution field
are shown in figure 2 where we plot the 0.5 spaced contours of 
the galaxy density contrast $\delta$  in the Supergalactic Plane. 
It is clear how NOG can constrain the shape and  dimensionality of 
high-amplitude, nearby structures as the so-called Supergalactic Plane. 
It is also clear from a first visual impression how much irregulars 
are the shapes of the major structures and  how much  rougly symmetric         
is the distribution  of high and low-density regions.

A full description of the density peaks and voids
characterizing the whole volume of the catalog will be presented in 
a forthcoming paper.

\begin{figure}
\plotfiddle{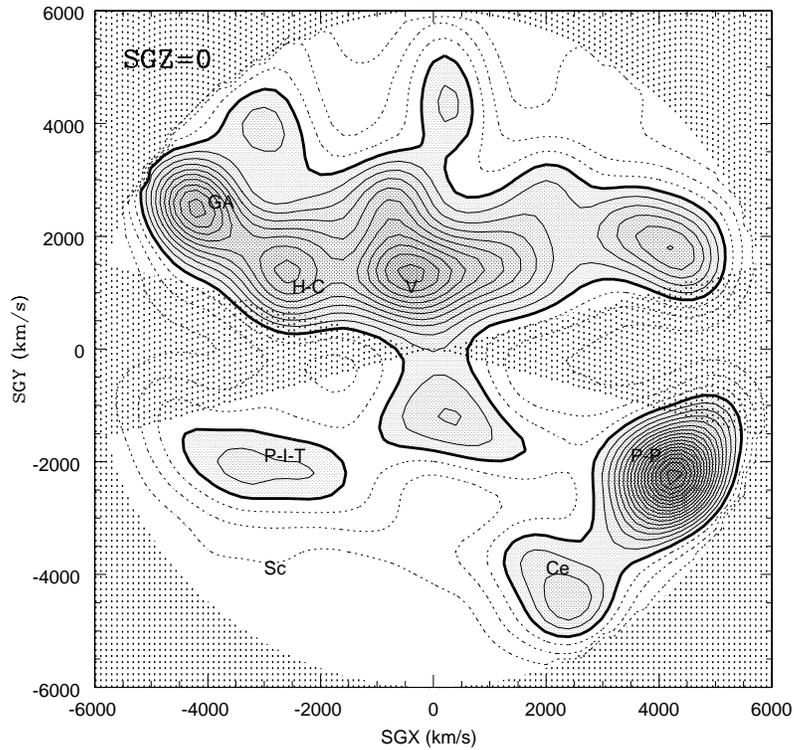}{8.5cm}{0}{53}{53}{-183}{-90}
\caption{The {\em real space} density field of NOG galaxies in the 
Supergalactic Plane. 
A Gaussian filter with an average smoothing length of 500 km/s has been 
applied.  Dashed contours represent negative values of $\delta$ 
i.e. underdense  regions with respect to the average density.
Some prominent structures dominating the local volume
such as  the Hydra-Centaurus-GA complex, Virgo, Perseus-Pisces, Cetus Wall 
and Sculptor void are clearly visible.
}
\end{figure}

\end{document}